\documentclass[11pt]{article}
\usepackage[margin=1in]{geometry}
\usepackage{graphics, array}
\usepackage{multirow}
\usepackage{harvard}
\usepackage{pslatex}
\usepackage{paralist}
\usepackage{amsmath, amsthm, amssymb} 

\begin{document}
\title{\sc Valuation Bounds of Tranche Options}
\date{Feb 4, 2010}
\author{Yadong Li and Ariye Shater 
\thanks{yadong.li@gmail.com, ariye.shater@barcap.com: the views expressed in 
this paper are the authors' own, and they do not necessarily reflect those 
of Barclays Capital. The authors thank Marco Naldi for many helpful comments
and discussions.} \\
Qauntitative Analytics, Barclays Capital}

\maketitle

\begin{abstract}

We performed a comprehensive analysis on the price bounds of CDO tranche 
options, and illustrated that the CDO tranche option prices can be effectively
bounded by the joint distribution of default time ($JDDT$) from a default
time copula. Systemic and idiosyncratic factors beyond the $JDDT$ only 
contribute a limited amount of pricing uncertainty. The price 
bounds of tranche option derived from a default time copula are often very 
narrow, especially for the senior part 
of the capital structure where there is the most market interests for 
tranche options. The tranche option bounds from a default time copula
can often be computed semi-analytically without Monte Carlo simulation,
therefore it is feasible and practical to price and risk manage senior 
CDO tranche options using the price bounds from a default time copula 
only.

CDO tranche option pricing is important in a number of practical situations 
such as counterparty, gap or liquidation risk; the methodology 
described in this paper can be very useful in the above described situations.
\end{abstract}

\section{Introduction}

The credit derivative market has experienced tremendous volatility since 
the beginning of the sub-prime and credit crisis. The standard credit index 
swaps and index tranches have become very important instruments for market 
participants to hedge or take positions on the overall credit quality 
and credit correlation. The index swaption market has become more active 
recently because of the increasing need to manage the volatility of market-wide 
credit movements. On the other hand, the index tranche option market
never  gained any traction despite the large realized volatility in the index tranches. 
The reasons are three fold: first, the index tranches are less liquid than the 
index swaps, secondly the standard index tranches can be viewed as an option 
on the index portfolio loss and it already provides leverages, therefore there is no 
need for investors to trade index tranche options in order to get leveraged 
exposure; thirdly, there is no standard model that can price and hedge index 
tranche options.  It remains a very challenging modelling problem to properly 
price CDO tranche options and the market participants generally lack the 
confidence in pricing and hedging the 
index tranche options. Despite the lack of interest to trade tranche options 
directly, it is very important to study the valuation of tranche options 
since they naturally arise from a number of common practical situations, for example in 
conterparty risk, gap risk or liquidation risk. 

Under current market conditions, it is almost impossible to price tranche options 
precisely because of the lack of relevant market observables. \cite{cyrus} suggested a 
method to compute the range bounds of tranche options from a default 
time copula. The key contribution of that paper is a scheme to compute the range 
bounds without the nested Monte Carlo simulation, which leads to easy calculations of 
the price bounds implied by a default time copula via a regular Monte Carlo simulation. 

A default time copula, by definition, only models the joint distribution of 
default time ($JDDT$), and it does not model any other factors. On the other 
hand, a bottom-up dynamic spread model attempts to model the joint distribution 
of default time and all the systemic and idiosyncratic factors that affect the 
spread dynamics. Dynamic factor model is the most common approach to build a 
bottom-up dynamic spread model, where the spread dynamics are driven by a few 
systemic factors that affects all the names and an idiosyncratic factor for 
each individual name. The idiosyncratic factors are easy to model because they 
are independent from other factors by definition, therefore the main task 
of building a dynamic factor model is to construct the joint distribution 
of default time and systematic factors ($JDDTSF$) which fully 
specifies the systemic dynamics.

The dynamic factor model is more difficult to build and calibrate than a default 
time copula since it needs to model more factors beyond the default time. 
The methodology prescribed by \cite{cyrus} can be very 
useful in practice if the resulting option price bounds from the default time
copula are narrow; as it allow us to price and risk manage tranche options 
without implementing a full dynamic factor model. \cite{cyrus} have shown that 
the price bounds of tranche options from a standard Gaussian Copula model are 
very narrow; however it is unclear if the price bounds would remain narrow in a more 
realistic situation where the default time copula has to be calibrated to the 
index tranche market across multiple maturities.

Recently, we suggested a very flexible dynamic correlation modelling framework 
in \cite{self}. A key finding of that paper is that the portfolio loss 
distribution and CDO tranche prices only depends on the joint distribution
of default indicators ($JDDI$); the modelling framework is more flexible than 
previous bottom-up models in the literature as it allows the $JDDT$ and $JDDTSF$ 
to change independently from the $JDDI$. With \cite{self}, once we calibrated 
the $JDDI$ to index tranche prices, we can easily construct different default time 
copula or dynamic factor models without changing the calibrated index tranche 
prices.  In this study, we used the \cite{self} model to construct different 
default time copulas from the same index tranche calibration, and we systematically 
study the price bounds of tranche options under these default time copulas. 

The paper is organized as follows: In section \ref{mkt} we first review some practical 
situations that involve the pricing of tranche options; then we review
the general derivation of the price bounds for tranche options in section 
\ref{bdd}; then we study the pricing bounds for European style tranche options 
in section \ref{fixedtr}; then we discuss tranche options with random triggers 
in section \ref{rndtr}. 

Even though we focus 
on tranche options in this paper, the methodology and conclusions are generally 
applicable to other types of multi-name credit options, such as options on NtD basket, 
and options on multiple tranches or CDO$^2$s.

\section{Practical Examples of Tranche Options\label{mkt}}
Tranche options naturally arise from a number of practical situations. We first 
define some terminology before reviewing these situations. Suppose there 
exists a probability space $(\Omega, \mathcal{F}_t, \mathbb{P})$ equipped 
with a risk-neutral probability measure $\mathbb{P}$. Consider a CDO tranche 
with a fixed set of payment date $\{t_i\}$ and a stream of cashflow $\{c_i\}$
 on the payment grid. The MTM of the tranche at time $t$ is 
 $ V_t = B_t \mathbb{E}[\sum_{t_i > t} \frac{c_i} {B_i} | \mathcal{F}_t]$, where $B_i$ is value
 of a money market account at $t_i$ that started with amount 1 at time 0. We further assume that 
 $B_i$ and $c_i$ are uncorrelated, therefore:
\begin{equation*}
 V_t = \sum_{t_i > t} d(t, t_i) \mathbb{E}[c_i | \mathcal{F}_t]
\end{equation*}
where $d(t, t_i) = \frac{B_t}{\mathbb{E}[B_{t_i} |\mathcal{F}_t]}$ 
is the risk free discount factor between $t$ and $t_i$.

\subsection{Counterparty Risk}
This is the classic case considered in \cite{cyrus}. Suppose a bank traded a 
tranche with a risky counterparty, if the counterparty default at time $t$, 
then the bank usually need to pay the full MTM ($V_t$) to the bankruptcy 
pool if the trade is to the counterparty's favor ($V_t < 0$), and the bank 
only recover a portion of the MTM if the trade's MTM is to the bank's 
favor ($V_t > 0$). Define $R$ as the recovery rate of the counterparty, 
then the bank could suffer a loss if the counterparty default, and the 
amount of the lose is $(1-R) \max[V_t, 0]$, which 
is a typical call option payoff. The counterparty effectively holds a call 
option to default and walk away from the remaining trade. The price of 
this call option to the counterparty is therefore:
\begin{equation*}
\textup{CP} = \mathbb{E}[ d(0, t) {\bf 1}_{\tau = t} \max[(1-R)V_t, 0] ]
\end{equation*}
The fair MTM price of the trade to the bank therefore has to be adjusted 
down to $V_0 - \textup{CP}$ if we price in the counterparty default risk. In reality,
the bank also holds an option to default, which can be priced similarly.

\subsection{Gap Risk}
Suppose a bank entered a tranche trade with a client, and the client
posted collateral in the amount of $C_0$ according to certain margin policies. 
Normally the collateral agreement allows the bank to make a margin call for 
additional collateral if the market moves against the client and the initial
margin is inadequate to cover the potential loss of the trade. 
Suppose $V_t$ is the MTM to the bank, and a margin call is made at time 
$t$, if the client does not post additional collateral within a certain time 
period $\delta$, which is typically a few days to two weeks after the margin call, 
the bank can seize the collateral and unwind the trade. A rationale client 
would choose not to post any additional collateral in 
the event of $C_0 < V_{t+\delta}$, therefore the client effectively holds 
a call option whose payoff is $\max(V_{t+\delta} - C_0, 0)$. Denote $\tau$ 
as the stopping time of the margin call, then the price of the call option 
to the client is:
\begin{equation*}
\textup{GAP} = \mathbb{E}[ d(0, t + \delta) {\bf 1}_{\tau = t} \max(V_{t+\delta} - C_0, 0) ]
\end{equation*}
The fair price of the instrument to the bank with this gap risk is therefore 
$V_0 - \textup{GAP}$. 

\subsection{Levered Super Senior Tranche}
Levered super senior (LSS) trade is a very popular trade for a client to 
take on leveraged risk on the senior part of the capital structure. In a typical 
LSS trade, the client sell protection on a super senior tranche $V_t$ to a bank
and the client only post collateral in the amount of $C_0$. The ratio between 
the notional amount of the $V_t$ and $C_0$ is the leverage factor. The LSS trade 
is different from the situation in the gap risk in that the bank can
only call for additional collateral if a pre-defined trigger event occurs. The 
trigger event can be portfolio loss reaching certain level, or tranche spreads
reaching certain level. Before the trigger event, the bank can't call additional 
collateral beyond $C_0$ even if the market moves against the client and the 
MTM of $V_t$ to the bank becomes greater than the collateral value $C_0$. After 
the trigger event, the bank usually is free to call 
additional collateral based on the MTM of the super senior tranche $V_t$. 

Since a rational client would not post any additional collateral if $C_0 < V_t$,
the bank only gets the smaller of $C_0$ and $V_t$ when trigger event occurs, 
therefore the value of the LSS trade to the bank is:
\begin{align*}
\textup{LSS} &= \mathbb{E}[ \sum_i d(0, t_i) {\bf 1}_{t_i<\tau} c_i + d(0, t){\bf 1}_{\tau = t} \min(V_{t},  C_0) ] \\
    &= \mathbb{E}[\sum_i d(0, t_i) {\bf 1}_{t_i<\tau} c_i + d(0, t) {\bf 1}_{\tau = t} (V_t - \max(V_{t} - C_0,  0)] \\
    &= \mathbb{E}[\sum_i d(0, t_i) {\bf 1}_{t_i<\tau} c_i + d(0,t){\bf 1}_{\tau = t} V_t] - \mathbb{E}[ d(0, t) {\bf 1}_{\tau = t}\max(V_{t} - C_0,  0)]
\end{align*}
where $c_i$ is the coupon payment for the super senior tranche, here we made 
the assumption that the trigger event always occurs before the super senior tranche 
suffers any real losses, which is almost always the case in practice since the 
trigger is put in to protect the bank thus it is designed to trigger far before 
the realized loss hits the tranche attachment. If the trigger is based on the 
portfolio loss, the first term can be computed from a default 
time copula. The second term is the value of a call option for the client to walk 
away from the trade when the trigger event occurs. 

The LSS trade is often mistakenly
modeled as a gap risk trade. Comparing the LSS trade with the gap risk, it is 
obvious that the LSS protection worths much less to the bank than in the case of 
the gap risk. Readers are referred to \cite{lss} for a very detailed discussion 
of the LSS.

\subsection{Liquidation Risk}
Suppose a client entered a funded credit-linked-note (CLN) trade referencing a 
tranche with a bank. At trade inception, the client deposits
the face amount of the CLN to a SPV, this principal may be 
invested in a risky asset $A_t$ for additional yield. The SPV then enters a 
unfunded swap contract with the bank to get exposure to the underlying 
tranche. We denote the MTM of the unfunded swap to the bank as $V_t$. The 
coupon payments from both the $A_t$ and $V_t$, netting of any fees to the 
bank, will be paid to the client. If either the underlying tranche gets 
impaired, or if the risky collateral $A_t$ defaults, both the $V_t$ and $A_t$ 
are liquidated; the client then receives the liquidation value of ($A_t - V_t$) 
if it is positive. Since the client never put additional money into the SPV 
besides the initial principal, the bank will suffer a loss in the event of 
the net liquidation value $A_t - V_t$ is negative. Define $\tau$ to be the 
stopping time of the liquidation event, then the client effectively holds a 
liquidation option whose payoff is $\max(V_t - A_t, 0)$ whose value to 
the client is:
\begin{equation*}
\textup{LIQ} = \mathbb{E}[ d(0, t) {\bf 1}_{\tau = t} \max(V_t - A_t, 0) ]
\end{equation*}
The fair value of the swap to the bank is therefore $V_0 - \textup{LIQ}$. The 
liquidation risk is similar to an exchange option between two assets $V_t$ 
and $A_t$.

There are several variations of the liquidation risk: in the 
famous (or infamous) mini-bond structure, $V_t$ is a first-to-default basket and $A_t$ 
is a synthetic CDO tranche. In a typical credit-linked-note, the $V_t$ can 
be a single name CDS or synthetic CDO tranche, and $A_t$ is very safe 
money market instruments. In a funding trade, the $V_t$ can be a CDO tranche or 
a single name CDS, and the $A_t$ is the term bond or funding of 
the bank.

\subsection{Callable Tranche}
Suppose a bank bought tranche protection from a client, and the trade's 
MTM to the bank is $V_t$. If the client is given the option to buy back 
the tranche protection at price $K$ under certain trigger event, then the 
client hold an option of $\max(V_t - K, 0)$ when the trigger event occurs. 
Denote $\tau$ as the stopping time of the trigger event, the client's option 
to call the tranche can be valued as:
\begin{equation*}
\textup{CAL} = \mathbb{E}[ d(0, t) {\bf 1}_{\tau = t} \max(V_t - K, 0) ]
\end{equation*}
The fair value of the swap to the bank is therefore $V_0 - \textup{CAL}$.

\section{Derivation of Price Bounds\label{bdd}}
\label{ublb}
All of examples in section \ref{mkt} reduces to the same 
problem of valuing the following call option where the exercise
time is random: 
\begin{equation}
\label{option}
C = \mathbb{E}[ d(0, t) {\bf 1}_{\tau = t} \max(V_t-K, 0) ]
\end{equation}
with $K$ as the strike price of the call. $V_t$ may involve multiple
tranches or assets as in the case of liquidation risk. This call option is very 
difficult to price because it depends on the MTM $V_t$ at a future time. 
Normally this types of problem requires nested Monte Carlo simulation because 
$V_t$ itself is an expectation of all future cashflows. It also requires 
a full dynamic model capable of generating 
spread levels at a future time on a simulated path, such a model is 
nearly impossible to calibrate given the lack of liquidity in the 
tranche option market. \cite{cyrus} offered an elegant solution to compute 
the range bounds of the option value $C$ just from the filtration generated 
by default events and recovery rates only (denoted as $\mathcal{D}_t$). 
Note that the full market filtration $\mathcal{F}_t$ also include other
systemic and idiosyncratic factors beyond $\mathcal{D}_t$, therefore 
$\mathcal{D}_t \subset \mathcal{F}_t$. We also have to assume that the trigger
event $\tau$ is adapted to $\mathcal{D}_t$, which excludes the spread
triggers. 

In this section, we review the derivation of the range bounds of (\ref{option}). 
The upperbound of (\ref{option}) is the same method as described in \cite{cyrus}, 
the lowerbound of (\ref{option}) given here is an improvement over the 
method in \cite{cyrus}, which first appeared in \cite{clex}. 
The key in the derivation is the Jensen's inequality which states: 
$g(\mathbb{E}[x]) \le \mathbb{E}[g(x)]$ if $g(x)$ is a convex function. 
In particular, since $\max(x - K, 0)$ is a convex function of $x$: 
$\max(\mathbb{E}[x] - K, 0) \le \mathbb{E}[\max(x - K, 0)]$.

Recall that $V_t = \mathbb{E}[\sum_{t_i > t} d(t, t_i) c_i | \mathcal{F}_t]$ 
where $c_i$ are the cashflows of the trade. $c_i$ is assumed to be adapted to 
$\mathcal{D}_{t_i}$, which is usually the case in practice, i.e., the 
cashflows of multi-name credit derivatives normally are only functions 
of realized default and recovery scenarios. The upper bound of $C$ can 
be derived as:
\begin{align}
\label{ub}
C &= \mathbb{E}[ d(0, t) {\bf 1}_{\tau = t} \max(V_t - K, 0) ]  \notag \\
&= \mathbb{E}[ d(0, t) {\bf 1}_{\tau = t} \max( \mathbb{E}[\sum_{t_i > t} d(t, t_i) c_i | \mathcal{F}_t] - K, 0) ]  &:&\mbox { expand $V_t$} \notag \\
&\le \mathbb{E}[ d(0, t) {\bf 1}_{\tau = t}  \mathbb{E}[\max(\sum_{t_i > t} d(t, t_i) c_i - K, 0) | \mathcal{F}_t] ] &:&\mbox{ Jensen's inequality} \\ 
&= \mathbb{E}[  \mathbb{E}[d(0, t) {\bf 1}_{\tau = t} \max(\sum_{t_i > t} d(t, t_i) c_i - K, 0) | \mathcal{F}_t] ] &:&\mbox{ ${\bf 1}_{\tau=t}$ is adapted to $\mathcal{F}_t$} \notag \\
&= \mathbb{E}[d(0, t) {\bf 1}_{\tau = t} \max(\sum_{t_i > t} d(t, t_i) c_i - K, 0)] &:&\mbox{ iterative expectation}  \notag
\end{align}
It is very straight-forward to compute the upper bound from a Monte Carlo simulation of default 
times and recovery rates since there is no nested simulations. Suppose $\mathcal{Y}_t$ is a 
sub filtration of $\mathcal{F}_t$ that includes the trigger event. The lower bound of $C$ can be 
derived as:
\begin{align}
\label{lb}
\notag
C &= \mathbb{E}[ d(0, t) {\bf 1}_{\tau = t} \max(V_t - K, 0) ] \\ \notag
&= \mathbb{E}[\mathbb{E}[ d(0, t) {\bf 1}_{\tau = t} \max(V_t - K, 0) | \mathcal{Y}_t] ] &:&\mbox{ iterative expectation} \\ \notag
&= \mathbb{E}[d(0, t){\bf 1}_{\tau = t}  \mathbb{E}[\max(V_t - K, 0) | \mathcal{Y}_t] ] &:&\mbox { ${\bf 1}_{\tau = t}$ is adapted to $\mathcal{Y}_t$} \\
&\ge  \mathbb{E}[d(0, t){\bf 1}_{\tau = t} \max(\mathbb{E}[ V_t - K | \mathcal{Y}_t], 0) ] &:&\mbox{ Jensen's inequality } \\ \notag
&=  \mathbb{E}[d(0, t){\bf 1}_{\tau = t} \max(\mathbb{E}[ V_t | \mathcal{Y}_t] - K, 0) ] &:&\mbox{ K is constant} \\ \notag
&=  \mathbb{E}[d(0, t){\bf 1}_{\tau = t} \max(\mathbb{E}[ \mathbb{E}[\sum_{t_i > t} d(t, t_i) c_i | \mathcal{F}_t] | \mathcal{Y}_t] - K, 0) ] &:&\mbox{ expand $V_t$ } \\ \notag
&=  \mathbb{E}[d(0, t){\bf 1}_{\tau = t} \max(\mathbb{E}[ \sum_{t_i > t} d(t, t_i) c_i| \mathcal{Y}_t] - K, 0) ] &:&\mbox{ iterative expectation }\end{align}
The term $\mathbb{E}[ \sum_{t_i > t} d(0, t_i) c_i | \mathcal{Y}_t]$ is 
the expected total value of future cashflow conditioned 
on the information in $\mathcal{Y}_t$. The choice of $\mathcal{Y}_t$ determines 
the quality of the lower bound, the more information in $\mathcal{Y}_t$ the higher
the lower bound is. In the limiting case of $\mathcal{Y}_t = \mathcal{F}_t$, the 
lower bound converges to the true value of the option. 

There is a very intuitive explanation of the upper and lower bounds of the 
option values. The MTM of the underlying tranche ($V_t$) is based on the information in the 
market filtration $\mathcal{F}_t$. In (\ref{ub}), the upper bound of option 
payoff ${\bf 1}_{\tau = t} \max(\sum_{t_i > t} d(0, t_i) c_i, 0)]$ corresponds 
to the option's value to an all-powerful deity who can perfectly foresee the future 
default events and recovery rates.  Therefore, at the time of the trigger event 
$\tau = t$, the deity will exercise the option based on the foreseeable future 
cashflow $\sum_{t_i > t} d(0, t_i) c_i$ instead of $V_t$. For example, the 
deity may exercise the option and buy protection on a CDO tranche even if the
MTM of the tranche is less than the strike price because he  
foresees that the future loss of the tranche will eventually exceed the future
value of the strike price. 
Therefore, the deity can extract more value from the option than its fair 
market value by exercising the option based on future information that is 
not part of $\mathcal{F}_t$. Therefore, the upper bound corresponds to the
option value with the divine power of perfect foresight.

The lower bound of the option payoff 
${\bf 1}_{\tau = t} \max(\mathbb{E}[ \sum_{t_i > t} d(0, t_i) c_i | \mathcal{Y}_t], 0)$ 
corresponds to an imprisoned investor who were only given the information in 
the sub-filtration $\mathcal{Y}_t \subset \mathcal{F}_t$. Therefore, he does not
observe the fair MTM value $V_t$, and he can only exercise the option based on
$\mathbb{E}[ V_t | \mathcal{Y}_t]$, which is an estimation (or best guess) of 
the MTM based on the available information $\mathcal{Y}_t$. This clearly results in 
suboptimal exercise of the option. Therefore, the lower bound corresponds to the 
option value with incomplete information. 

Even with a full dynamic model, one could obtain the lower bound of the option
instead of the true value if the numerical methods of the option pricing is built 
on a reduced filtration, which is often the case with the lattice methods. 
For example, in the pioneering work by \cite{ml}, the tranche options 
are priced by building a lattice on a reduced filtration with low dimensionality, 
which results in a lower bound of the option in the strict sense even though the
authors were trying to obtain the true value of the option. 

In the most generic form, the upper bound and lower bound can be computed 
from the Monte Carlo simulation of a default time copula. We don't need to 
model or simulate any future spreads in order to compute the price bounds if 
we choose $\mathcal{Y}_t \subset \mathcal{D}_t$. The upper bound can be computed 
directly from the simulated default time and recoveries of all the underlying 
credits, and the lower bound requires a least square Monte Carlo simulation 
as in \cite{ls} that regresses the value of the future cash flows to the state
variables in $\mathcal{Y}_t$. Semi-analytical solution to the option 
bounds can be obtained if the trigger event is deterministic in time (ie, vanilla 
European option whose holder can exercise
at a deterministic future time), or if the trigger event is the default event of
a single credit, such as the case in the counterparty risk. We'll analyze these
special cases in the following sections. 

The bounds derived from the default time copula offers great insights on the 
tranche option pricing. If the bounds are very narrow,  the option values are mainly 
determined by the $JDDT$; if the bounds are wide, then the option values are primarily 
determined by other systemic or idiosyncratic factors beyond the default time. In this 
paper, we'll try to understand what is the main driver of the tranche option prices.

The price bounds for put option can be obtained via the put-call parity. 
The methodology to obtain the lower and upper bound of an option payoff is 
very general, it applies to any multi-name credit options, such as CDO tranche 
option, NTD basket option or the multiple asset options as discussed in
liquidation risk.

\section{European Tranche Options\label{fixedtr}}
In this section, we consider the vanilla European tranche options which 
can be exercised at a pre-determined time. The tranche options with random 
trigger event will be discussed in the next section. 

For simplicity, we 
consider the tranche loss option instead of the more general case of tranche 
option. An European tranche loss option is a hypothetical instrument that gives 
the buyer the right (not an obligation) to pay a fixed amount $K$ at the exercise 
time $t$ in order to receive a payoff equal to the total realized loss of 
a tranche at time $T$. Regular tranche option reduces to the tranche loss option if the 
tranche has no running coupon and if we assume the protection payments are 
all made at maturity instead of the time of default. The price bounds of 
tranche loss option with deterministic time trigger can be computed 
without Monte Carlo simulation. Since 
the main driver of a tranche's value is its expected tranche loss at maturity, the 
conclusions drew from the analysis on the tranche loss option 
applies to the more general cases of tranche option with running coupons and
immediate protection settlement. Furthermore, the regular tranche option can be approximated 
using tranche loss option, please see Appendix \ref{a_tr} for a more detailed
discussion of the approximation.

Then the PV of the tranche loss option can be written as:
\begin{align}
\label{tlob}
C &=  d(0, t) \mathbb{E}[\max( d(t, T) \mathbb{E}[ L_T(A, D) | \mathcal{F}_t] - K, 0) ] \notag \\
	&=  d(0, T) \mathbb{E}[\max(\mathbb{E}[ L_T(A, D) | \mathcal{F}_t] - \frac{K}{d(t,T)}, 0) ]
\end{align}

We use $A$ and $D$ to denote the tranche's attachment and detachment levels.
In the subsequent analysis, we drop all the deterministic discount factors to 
simplify the exposition, with the understanding that the valuation 
bounds and strikes need to be adjusted with those deterministic discount 
factors in \eqref{tlob}. The $L_T(A, D)$ in \eqref{tlob} is the expected 
tranche loss (ETL) and $L_T$ is the portfolio loss at tranche maturity $T$: 
\begin{equation*}
L_T(A, D) = \min(\max(L_T-A, 0), D-A)
\end{equation*}	

A simple expression for the upperbound can be derived from (\ref{ub}):
\begin{align}
\label{ubloss}
C &\le  \mathbb{E}[\max(L_T(A, D) - K, 0)] \notag \\
	&= \mathbb{E}[\max(\min(\max(L_T-A, 0), D-A) - K, 0)))] \notag \\
	&= \mathbb{E}[\min(\max(L_T-(A+K), 0), D-(A+K))] \\ \notag
	&= \mathbb{E}[L_T(A+K, D)]
\end{align}
Therefore, the upper bound of the tranche loss option is just the ETL 
of an $A+K$ to $D$ tranche. This relationship does not 
hold if the tranche has a non-zero running coupon or if the protection 
payment is not made at the end of tranche maturity. However, since the 
impact of running coupon and the discounting of protection payments is 
limited in the tranche pricing, we can still use the PV of a $A+K$ 
to $D$ tranche as an approximation to the upper bound of a regular tranche 
option. This is a very handy relationship in practice. A more accurate 
upper bound for the regular tranche option can be obtained using the
approximation in Appendix \ref{a_tr}. 

The upper bound of the tranche loss option only depends on the terminal 
loss distribution, therefore, it is not model dependent as long as all 
the models are calibrated to the same loss distribution. For example,
we can compute the upper bound of a tranche option even from a
base correlation model.

The lower bound of the tranche loss option can be written as:
 \begin{equation}
\label{lbloss}
C \ge  \mathbb{E}[\max(\mathbb{E}[L_T(A, D)|\mathcal{Y}_t] - K, 0)] 
\end{equation}
where $\mathcal{Y}_t$ is a sub-filtration of the market filtration 
$\mathcal{F}_t$. The lower bound is generally model dependent through 
the conditional expectation $\mathbb{E}[L_T(A, D)|\mathcal{Y}_t]$ but 
we can derive a naive model-independent lower bound by Jensen's inequality: 
 \begin{align}
 \label{naive_lb}
C &=  \mathbb{E}[\max(\mathbb{E}[L_T(A, D)|\mathcal{F}_t] - K, 0)] \notag \\
&\ge \max(\mathbb{E}[\mathbb{E}[L_T(A, D)|\mathcal{F}_t] - K], 0) \\
&=  \max(\mathbb{E}[L_T(A, D)] - K, 0) \notag
\end{align}
The lower bound in \eqref{naive_lb} corresponds to an exercising strategy that 
the option holder always exercises the option if the expected tranche 
loss based on information at $t=0$ is more than the strike price $K$, aka, 
if the option is ``in-the-money'' at $t=0$. \cite{lss} pointed out that 
a digital tranche is the upper bound (``super-hedge'') of a loss-trigger
LSS trade. The digital tranche is equivalent to the naive lower 
bound \eqref{naive_lb} in the context of the LSS. 

To get more precise lower bound, we have to choose the sub-filtration 
$\mathcal{Y}_t$ with more information. We used the 
CDX-IG9 index tranches and market data on Jul 21st, 2009 for 
this study. We calibrated the model described in \cite{self} 
to the market data, and Table \ref{etl} showed the expected tranche loss 
from the calibrated model. Note that all the ETLs are normalized to their 
tranche notionals, so are the option values in the rest of this 
document\footnote{We still refer the CDX-IG9 tranches using their original strikes even 
though the actual calculations were using the adjusted strikes which 
take into account the three defaulted names in the portfolio.}.

\begin{table}
\caption{CDX-IG9 Expected Tranche Loss}
\label{etl}
\center

\footnotesize

\begin{tabular}{|c|rrrr|}
\hline
Tranches & 3Y & 5Y & 7Y & 10Y\\
\hline
0-3\% & 54.12\% & 80.19\% & 86.76\% & 91.12\%\\
3-7\% & 17.03\% & 42.64\% & 55.16\% & 66.18\%\\
7-10\% & 5.36\% & 20.09\% & 33.98\% & 48.18\%\\
10-15\% & 1.35\% & 8.17\% & 15.82\% & 23.34\%\\
15-30\% & 0.76\% & 2.29\% & 4.81\% & 7.95\%\\
30-60\% & 0.49\% & 1.62\% & 3.40\% & 5.31\%\\
60-100\% & 0.02\% & 0.42\% & 0.95\% & 1.54\%\\
\hline
\end{tabular}

\end{table}

We first consider IG9 tranche loss options that expires at 3Y 
(maturity: Dec. 20, 2010) for the expected loss of a 5Y (maturity: Dec. 20, 2012) 
tranche, we define the at-the-money (ATM) strike to be the expected tranche loss 
$K^{ATM} = \mathbb{E}[L_T(A, D)]$. In this study, we also computed the 
price bounds for in-the-money(ITM) and out-of-the-money 
(OTM) tranche loss options. In the following examples, the ITM strike is half of 
ETL, and the OTM strike is twice of the ETL: 
$K^{ITM} = \frac{1}{2}\mathbb{E}[L_T(A, D)]$ and $K^{OTM} = 2\mathbb{E}[L_T(A, D)]$.
Table \ref{upper} showed the upper bounds computed from the calibrated bottom-up 
model according to \eqref{ubloss}. 
\begin{table}
\caption{Upper Bounds of 3Y-5Y Tranche Loss Option \label{upper}}

\center
\footnotesize

\begin{tabular}{|c|rrr|}
\hline
CDX-IG9 & \multicolumn{3}{|c|}{Upper Bounds}\\
\cline{2-4}
 Tranches & ITM & ATM & OTM\\
\hline
0-3\% & 43.33\% & 12.73\% & 0.00\%\\
3-7\% & 30.71\% & 20.57\% & 4.44\%\\
7-10\% & 17.35\% & 14.79\% & 10.21\%\\
10-15\% & 7.63\% & 7.11\% & 6.14\%\\
15-30\% & 2.24\% & 2.19\% & 2.10\%\\
30-60\% & 1.61\% & 1.59\% & 1.56\%\\
60-100\% & 0.42\% & 0.41\% & 0.41\%\\
\hline
\end{tabular}

\end{table}

We now focus on the lower bounds which depends on the choice of the 
sub-filtration $\mathcal{Y}_t$.

\subsection{Lower Bounds from Top-down Models}
The minimum sub-filtration that can price tranche loss option 
consistently is the filtration generated by the portfolio loss process, 
we denote it as $\mathcal{L}_t$. Note that $\mathcal{L}_t$ does not 
contain any single name information and typical top-down models are 
built on the $\mathcal{L}_t$ filtration. 

The $\mathcal{L}_t$ does impose a more precise lower bounds of the 
tranche loss option than (\ref{naive_lb}). To 
illustrate this, we take a discrete sample of the initial loss 
distribution, and built two different Markov chains on the loss 
distribution: co-monotonic Markov chain and maximum entropy Markov 
chain. The details of how to build these Markov chains can be found 
in \cite{losslinker}. Once we have a Markov chain on the loss 
transition, we can then compute the lower bound in (\ref{lbloss}) by 
conditioning on the portfolio loss $L_t$. Since the conditioning
is only on a scaler variable, the lower bound can be easily computed 
from the Markov Chain without using Monte Carlo simulation. The lower 
bounds from the two different Markov chains are shown in the table \ref{bdl}: 
the co-monotonic Markov chain implies a much higher lower bound 
than the maximum entropy Markov chain. The OTM option on the 0-3\% 
equity tranche has a value of 0 since the OTM strike is more than 
the tranche notional. The 60-100\% tranche's lower bounds with 
co-monotonic Markov chain are slightly higher than the upper bounds in 
Table \ref{upper}, which is caused by the inaccuracies of the discrete 
sampling of the loss distribution. 

An interesting question is: what is the lower bound if we only know 
the loss distributions but not the Markov chain of loss transition?  
This bound is of special interest 
because it is not model dependent, and it is the lowest lower bound among 
all admissible Markov chains by the loss distribution. We denote this 
lowest lower bound as LLB($\mathcal{L}_t$). Finding 
the LLB($\mathcal{L}_t$) among all possible Markov chains can be formulated 
as a nonlinear optimization problem (see the Appendix \ref{a_opt}), which 
can be solved using a standard non-linear optimizer. The column ``LLB'' 
in table \ref{bdl} is the lowest lower bound obtained from the nonlinear 
optimization. Note that the optimizations to find the LLBs for different 
tranches are run separately, therefore the tranche loss options from 
different tranches can't be at their LLB($\mathcal{L}_t$) simultaneously. 
For example, if the tranche loss option 
for the 0-3\% tranche is priced at its LLB($\mathcal{L}_t$), then the 3-7\% 
tranche loss option price has to be greater than its LLB($\mathcal{L}_t$) since 
the Markov chain that produces the LLB($\mathcal{L}_t$) for 0-3\% tranche 
is generally not the same Markov chain that produces the LLB($\mathcal{L}_t$) 
of 3-7\%. Though still crude, the LLB($\mathcal{L}_t$) from loss distribution 
is much more precise than the naive lower bound in (\ref{naive_lb}), 
which are zeros for all the ATM or OTM options.

\begin{table}
\caption{Lower Bounds of 3Y-5Y Option from $\mathcal{L}_t$ (Top-down) \label{bdl}} 
\center

\footnotesize

\begin{tabular}{|c|rrr|rrr|rrr|}
\hline
CDX-IG9 & \multicolumn{3}{|c|}{ITM Lower Bounds} & \multicolumn{3}{|c|}{ATM Lower Bounds} & \multicolumn{3}{|c|}{OTM Lower Bounds}\\
\cline{2-10}
Tranches & Co-mo & Max-E & LLB & Co-mo & Max-E & LLB & Co-mo & Max-E & LLB\\
\hline
0-3\% & 43.09\% & 39.97\% & 39.97\% & 12.50\% & 7.91\% & 6.40\% & 0.00\% & 0.00\% & 0.00\%\\
3-7\% & 30.32\% & 22.34\% & 21.40\% & 19.75\% & 11.74\% & 7.55\% & 4.18\% & 1.82\% & 1.54\%\\
7-10\% & 17.01\% & 11.05\% & 9.82\% & 14.39\% & 7.53\% & 3.67\% & 9.35\% & 3.99\% & 2.28\%\\
10-15\% & 7.63\% & 4.88\% & 4.20\% & 7.13\% & 3.64\% & 1.07\% & 6.28\% & 2.38\% & 0.94\%\\
15-30\% & 2.19\% & 1.48\% & 1.12\% & 2.11\% & 1.24\% & 0.68\% & 2.02\% & 1.00\% & 0.66\%\\
30-60\% & 1.60\% & 1.11\% & 0.81\% & 1.57\% & 0.97\% & 0.41\% & 1.52\% & 0.83\% & 0.40\%\\
60-100\% & 0.48\% & 0.35\% & 0.28\% & 0.48\% & 0.30\% & 0.07\% & 0.47\% & 0.26\% & 0.03\%\\
\hline
\end{tabular}
\end{table}

In table \ref{bdl}, the LLB($\mathcal{L}_t$) are greater than 0 for all 
tranches even in the case of OTM options. This is a unique feature of 
the CDO tranche option. In other asset classes, the OTM option values can 
be very close to 0 if the volatility of the underlying asset becomes very low. However, 
even the OTM tranche option always have certain minimum value regardless
of the tranche spread volatility. The reason is that the dynamics of 
the portfolio loss process has to be consistent 
with the initial loss distribution at time $t=0$, which imposes a minimum 
level of portfolio loss volatility. For example, the volatility of portfolio 
loss process cannot be 0 since a deterministic portfolio loss process clearly 
violates the initial loss distribution at $t=0$. 

The value of a tranche option depends on the full loss distribution hence
it is important to model tranche options on the same 
underlying portfolio across capital structure  as inter-dependent instruments.
 The simple approach of modeling tranche options as separate derivative instruments on  
individual tranches, as suggested by \cite{hwoption}, is not adequate.
 \cite{hwoption} attempted to model tranche options using a 
similar approach to the Libor market model in the interest rates world, which 
could produce inconsistent prices with the underlying tranche prices. 
For example, there is no restriction on the volatility parameter of the 
forward tranche spread in the \cite{hwoption} approach and we could 
produce arbitrage-able tranche option prices out of the range bound 
from \eqref{ub} and \eqref{lb} by choosing the volatility 
parameter. 

\subsection{Lower Bound from the (Li 2009) Model}
More precise lower bound can be obtained if the $\mathcal{Y}_t$ 
in \eqref{lb} also includes single name information. To study the effects
of single name information, we used the model described in \cite{self}. The
\cite{self} model is a one-factor bottom-up dynamic model where the systemic 
factor is modeled by an increasing process $X_t$. Under the \cite{self} model, 
the marginal distribution of $X_t$ determines the $JDDI$; the Markov chain 
on $X_t$ determines the $JDDT$, therefore each different Markov chain on $X_t$
defines a different default time copula. The marginal distribution of $X_t$
can be calibrated to index tranche prices, afterwards, we can construct different
default time copulas by constructing different Markov chain to the marginal
distribution of $X_t$. These different default time copula produces different 
$JDDT$ but identical $JDDI$ and tranche prices by construction. 

Assume $\mathcal{S}_t$ is a a filtration generated by 
the common factor process $X_t$ and the single name default and recovery. Of
course $\mathcal{S}_t \subset \mathcal{F}_t$ since $\mathcal{F}_t$ includes
other systemic and idiosyncratic factors beyond $X_t$. In the numerical implementation, 
the lower bound from $\mathcal{S}_t$ are 
computed by only conditioning on the value of the common factor $X_t$ 
but not the loss $L_t$, which allow us to compute the lower bound with a 
semi-analytical method pioneered by \cite{semianalytical}. Ignoring the realized loss $L_t$
results in slightly worse (or lower) lower bounds because we are not using the full information
available in $\mathcal{S}_t$, but it is a good trade off since it allows
us to use the semi-analytical pricing method for much faster calculation
of the lower bound. Since the $X_t$ and the loss $L_t$ are highly correlated 
under the one-factor model, the degradation of the lower bound quality is 
expected to be small by excluding the realized losses in the conditioning. 

In this example, we built a co-monotonic Markov chain and a maximum entropy 
Markov chain on $X_t$, and the resulting lower bounds from $\mathcal{S}_t$ and 
these two default time copula are show in table \ref{bd1f}. We can also find the 
lowest lower bound of all possible Markov chains of $X_t$ that preserves the
$JDDI$ and tranche prices. We denote the lowest lower bound based on the 
sub-filtration $\mathcal{S}_t$ as LLB($\mathcal{S}_t$) (shown in the 
column ``LLB'' of table \ref{bd1f}). The LLB($\mathcal{S}_t$) 
is much higher than the LLB($\mathcal{L}_t$) because it is constrained 
by the additional single name information and the conditional independent 
correlation structure in the \cite{self} model. 

Comparing table \ref{bdl} and \ref{bd1f}, it is interesting to note that 
the lower bounds from $\mathcal{L}_t$ can vary at a much wider range than 
the lower bound from $\mathcal{S}_t$. This can be explained by the fact 
that not all the loss transitions are admissible under a factor model 
with conditional independence. 
For example, the co-monotonic Markov chain built on the loss process 
have many deterministic transitions, such as: if the tranche loss is 
2\% at 3Y, then the loss will be 4\% at 5Y with probability 1. The 
existence of such fully deterministic transition is a property of the 
co-monotonic Markov chain. Though admissible under a contagion model, 
the deterministic loss transition
is incompatible with a conditional independent factor model where the 
$L_T$ conditioned on $\mathcal{F}_t$ can never be fully deterministic 
except for the degenerated case. Therefore, adding single name information 
and a conditional independent correlation structure further restricts the 
set of admissible loss transitions, thus imposing a narrower range on 
the lower bounds. 

\cite{lando} have shown that the conditional independent assumption cannot 
be rejected from either the individual case studies or the statistical tests 
of the historical default events. None of the historical default events
so far are caused by contagion in the strict sense that one company's default
event directly caused another company to default. Contagion models also have some 
undesirable properties as shown in \cite{seb2} that make it difficult to 
use in practice. Therefore, 
conditional independent factor model remains the most practical and efficient 
approach to include single name information. In practice, we have to 
adopt the lower bounds from the conditional independent model since
it is the only feasible approach to price and manage both the vanilla 
CDO tranches and exotic instruments like tranche options. 

\begin{table}
\caption{Lower Bounds of 3Y-5Y Option from $\mathcal{S}_t$ \label{bd1f}}
\center

\footnotesize

\begin{tabular}{|c|rrr|rrr|rrr|}
\hline
CDX-IG9 & \multicolumn{3}{|c|}{ITM Lower Bounds} & \multicolumn{3}{|c|}{ATM Lower Bounds} & \multicolumn{3}{|c|}{OTM Lower Bounds}\\
\cline{2-10}
 Tranches & Co-mo & Max-E & LLB & Co-mo & Max-E & LLB & Co-mo & Max-E & LLB\\
\hline
0-3\% & 41.07\% & 40.93\% & 40.05\% & 10.47\% & 9.53\% & 8.46\% & 0.00\% & 0.00\% & 0.00\%\\
3-7\% & 29.15\% & 26.63\% & 22.07\% & 19.13\% & 15.45\% & 11.96\% & 3.19\% & 2.57\% & 2.47\%\\
7-10\% & 16.26\% & 14.48\% & 13.12\% & 13.20\% & 11.74\% & 11.20\% & 8.12\% & 7.85\% & 7.58\%\\
10-15\% & 7.32\% & 7.26\% & 6.42\% & 6.70\% & 6.64\% & 5.79\% & 5.54\% & 5.47\% & 4.85\%\\
15-30\% & 2.19\% & 2.14\% & 1.23\% & 2.12\% & 2.04\% & 1.16\% & 1.99\% & 1.90\% & 1.09\%\\
30-60\% & 1.60\% & 1.53\% & 0.82\% & 1.57\% & 1.47\% & 0.65\% & 1.52\% & 1.37\% & 0.73\%\\
60-100\% & 0.41\% & 0.39\% & 0.20\% & 0.41\% & 0.38\% & 0.12\% & 0.41\% & 0.35\% & 0.14\%\\
\hline
\end{tabular}

\end{table}

\subsection{Systemic vs. Idiosyncratic Dynamics}

Furthermore, we can quantify how much uncertainty of the option value is due 
to systemic dynamics vs. idiosyncratic dynamics under the \cite{self} model. 
Suppose we have a filtration $\mathcal{U}_t$ which include $\mathcal{S}_t$ 
and $X_T$, i.e., this filtration correspond to a less powerful deity 
(comparing to the all-powerful deity that gives the upper bound) who can 
only foresee the future value of the common factor, but not the idiosyncratic 
default events. The remaining uncertainty between the lower bound from $\mathcal{U}_t$ 
and the upper bound has to be 
caused by idiosyncratic dynamics. Therefore, the option bounds from 
$\mathcal{U}_t$ gives a way to gauge the pricing uncertainty purely due to 
the idiosyncratic dynamics. Table \ref{lbsys} showed the lower 
bounds calculated from $\mathcal{U}_t$. By comparing to the upper bounds in 
Table \ref{ub},  it is obvious that option pricing uncertainty due to 
idiosyncratic dynamics is very limited. The idiosyncratic dynamics only 
contributes a small amount of uncertainty to junior tranche options, and it 
has almost no contributions to the senior tranche options. Therefore, we can 
safely ignore the idiosyncratic spread dynamics if we are mainly dealing 
with the senior tranche options. 

\begin{table}
\caption{Lower Bounds of 3Y-5Y Option from $\mathcal{U}_t$ (Perfect Foresight) \label{lbsys}}
\center

\footnotesize
\begin{tabular}{|c|rrr|}
\hline
CDX-IG9 & \multicolumn{3}{|c|}{Lower Bounds}\\
\cline{2-4}
 Tranches & ITM & ATM & OTM\\
\hline
0-3\% & 41.14\% & 10.73\% & 0.00\%\\
3-7\% & 29.47\% & 19.15\% & 3.28\%\\
7-10\% & 16.37\% & 13.38\% & 8.18\%\\
10-15\% & 7.36\% & 6.70\% & 5.69\%\\
15-30\% & 2.20\% & 2.13\% & 2.04\%\\
30-60\% & 1.60\% & 1.59\% & 1.56\%\\
60-100\% & 0.42\% & 0.41\% & 0.41\%\\
\hline
\end{tabular}

\end{table}

\begin{table}
\caption{Price Bounds of 5Y-10Y Tranche Loss Option \label{bd10y}}
\center

\normalsize
\underline{In-the-Money Option}
\footnotesize

\vspace{.25cm}

\begin{tabular}{|c|rrrrr|r|}
\hline
CDX-IG9 & \multicolumn{5}{|c|}{Lower Bounds} & Upper\\
\cline{2-6}
 Tranches & LLB($\mathcal{L}_t$) & Max-E $\mathcal{L}_t$ & LLB($\mathcal{S}_t$) & Max-E $\mathcal{S}_t$ & $\mathcal{U}_t$ & Bound \\
\hline
0-3\% & 45.46\% & 45.58\% & 45.37\% & 45.56\% & 45.40\% & 46.97\%\\
3-7\% & 33.19\% & 35.75\% & 35.76\% & 39.64\% & 41.28\% & 42.12\%\\
7-10\% & 23.23\% & 26.75\% & 25.15\% & 30.90\% & 33.13\% & 34.66\%\\
10-15\% & 11.94\% & 14.25\% & 11.81\% & 16.47\% & 18.07\% & 19.29\%\\
15-30\% & 3.91\% & 5.08\% & 3.96\% & 6.58\% & 7.22\% & 7.42\%\\
30-60\% & 2.61\% & 3.49\% & 2.62\% & 4.60\% & 5.14\% & 5.14\%\\
60-100\% & 0.88\% & 1.16\% & 0.76\% & 1.37\% & 1.50\% & 1.50\%\\
\hline
\end{tabular}

\vspace{.5cm}

\normalsize
\underline{At-the-Money Option}
\footnotesize

\vspace{.25cm}

\begin{tabular}{|c|rrrrr|r|}
\hline
CDX-IG9 & \multicolumn{5}{|c|}{Lower Bounds} & Upper\\
\cline{2-6}
Tranches & LLB($\mathcal{L}_t$) & Max-E $\mathcal{L}_t$ & LLB($\mathcal{S}_t$) & Max-E $\mathcal{S}_t$ & $\mathcal{U}_t$ & Bounds \\
\hline
0-3\% & 5.61\% & 5.76\% & 4.96\% & 5.62\% & 5.62\% & 7.09\%\\
3-7\% & 12.91\% & 14.92\% & 15.80\% & 17.98\% & 19.99\% & 20.53\%\\
7-10\% & 8.16\% & 13.70\% & 11.82\% & 16.85\% & 18.49\% & 22.23\%\\
10-15\% & 5.60\% & 8.87\% & 9.27\% & 12.06\% & 14.33\% & 15.69\%\\
15-30\% & 1.98\% & 3.94\% & 3.49\% & 5.96\% & 6.75\% & 6.93\%\\
30-60\% & 1.52\% & 2.91\% & 1.51\% & 4.27\% & 4.97\% & 4.98\%\\
60-100\% & 0.46\% & 1.01\% & 0.56\% & 1.27\% & 1.47\% & 1.47\%\\
\hline
\end{tabular}

\vspace{.5cm}

\normalsize
\underline{Out-of-the-Money Option}
\footnotesize

\vspace{.25cm}

\begin{tabular}{|c|rrrrr|r|}
\hline
CDX-IG9 & \multicolumn{5}{|c|}{Lower Bounds} & Upper\\
\cline{2-6}
Tranches & LLB($\mathcal{L}_t$) & Max-E $\mathcal{L}_t$ & LLB($\mathcal{S}_t$) & Max-E $\mathcal{S}_t$ & $\mathcal{U}_t$ & Bound \\
\hline
0-3\% & 0.00\% & 0.00\% & 0.00\% & 0.00\% & 0.00\% & 0.00\%\\
3-7\% & 0.00\% & 0.00\% & 0.00\% & 0.00\% & 0.00\% & 0.00\%\\
7-10\% & 0.53\% & 0.53\% & 0.52\% & 0.50\% & 0.80\% & 1.36\%\\
10-15\% & 3.23\% & 4.57\% & 5.57\% & 6.47\% & 8.15\% & 9.65\%\\
15-30\% & 1.69\% & 2.98\% & 2.90\% & 5.00\% & 5.92\% & 6.08\%\\
30-60\% & 1.43\% & 2.33\% & 1.41\% & 3.70\% & 4.65\% & 4.66\%\\
60-100\% & 0.44\% & 0.86\% & 0.54\% & 1.15\% & 1.41\% & 1.41\%\\
\hline
\end{tabular}

\end{table}

\subsection{Long-dated Options}
The upper and lower bounds of a 5Y to 10Y tranche loss option are also computed in
Table \ref{bd10y}. In general, the price bounds of the 5Y-10Y options exhibit very 
similar features as the 3Y-5Y options. The 5Y-10Y option showed a wider range 
between upper and lower bound than the 3Y-5Y option, which is not surprising 
since the long-dated option is expected to have more pricing uncertainties. 

\subsection{Choice of Markov Chains}
In Table \ref{lbsys},  the lower bounds from $\mathcal{U}_t$ is only slightly higher than the 
lower bound from co-monotonic Markov chain because the co-monotonic Markov chain 
is very close to having perfect foresight as the common factors at the two 
maturities are mapped sequentially by their distribution quantiles. If the 
common factor is specified as a continuous distribution, the co-monotonic 
Markov chain will produce the exact same lower bound as those from $\mathcal{U}_t$.
Therefore, it is arguable that the co-monotonic Markov chain is not realistic due 
to the collapsed uncertainty of future common factor distribution. 

As noted by many previous authors, eg \cite{andersenfactor} and \cite{skarke}, 
the classic Gaussian Copula implies very 
unrealistic spread dynamics. This price bound analysis of tranche options offers 
yet another interesting view on the Gaussian Copula: the classic Gaussian 
Copula model is a degenerated co-monotonic 
Markov chain across time where the common factor distributions remains unchanged 
(Gaussian). Therefore, the classic Gaussian copula suffers from the same problem 
of vanishing common factor uncertainties as the co-monotonic Markov 
chain. Co-monotonic Markov chains, including Gaussian Copula, will overvalue the 
tranche options because of the perfect foresight of the future market factor 
realizations.

In comparison, the maximum entropy Markov chain is a much better choice since it 
has the advantage of keeping the least amount of information in the system, and 
the uncertainty of the future common factor is the largest among all possible
Markov chains (because the information entropy is maximized). Ideally we should 
calibrate the Markov chain using market information; however it is impossible to
do so under the current market condition because there is no relevant
and reliable market observables on the transition of future loss process. Given
the lack of market information, we argue 
that the maximum entropy Markov chain is the most natural choice since 
it corresponds to the state of no relevant information. Table \ref{bdl} 
to \ref{bd10y} have shown that the lower bounds become more precise with larger 
sub-filtrations, and the price bounds imposed by $\mathcal{S}_t$ and the maximum entropy 
Markov chains are still quite tight, especially for the senior tranches. Therefore, the 
dynamics of other systemic and idiosyncratic factors beyond $\mathcal{S}_t$ only 
have limited contribution to the pricing uncertainty of tranche options. 

\section{Tranche Options with Random Triggers\label{rndtr}}
As discussed in section \ref{mkt}, the trigger event itself for a tranche
option can be a random event. In general, the tranche options with random triggers 
require Monte Carlo simulation to compute its upper and lower bounds. However, if 
the trigger event is the default of a single credit, we can still treat it 
semi-analytically without Monte Carlo simulation by taking advantage of the 
conditional independence. 

\begin{table}
\caption{3Y-5Y ATM Options with Single Default Event Trigger\label{sntrig}}
\center

\underline{Trigger Default Prob = 5\%}

\footnotesize

\vspace{.25cm}

\begin{tabular}{|c|rr|rr|rr|}
\hline
CDX-IG9 & \multicolumn{2}{|c|}{Independent} & \multicolumn{2}{|c|}{Less Correlated} & \multicolumn{2}{|c|}{More Correlated}\\
\cline{2-7}
Tranches & LB & UB & LB & UB & LB & UB\\
\hline
0-3\% & 0.48\% & 0.64\% & 0.69\% & 0.73\% & 0.86\% & 0.87\%\\
3-7\% & 0.77\% & 1.03\% & 1.57\% & 1.69\% & 2.14\% & 2.22\%\\
7-10\% & 0.59\% & 0.74\% & 1.69\% & 1.75\% & 2.41\% & 2.46\%\\
10-15\% & 0.33\% & 0.36\% & 1.37\% & 1.37\% & 1.89\% & 1.90\%\\
15-30\% & 0.10\% & 0.11\% & 0.89\% & 0.90\% & 1.09\% & 1.11\%\\
30-60\% & 0.07\% & 0.08\% & 0.80\% & 0.81\% & 0.94\% & 0.96\%\\
60-100\% & 0.02\% & 0.02\% & 0.22\% & 0.22\% & 0.25\% & 0.26\%\\
\hline
\end{tabular}

\vspace{.5cm}
\normalsize
\underline{Trigger Default Prob = 30\%}
\footnotesize

\vspace{.25cm}

\begin{tabular}{|c|rr|rr|rr|}
\hline
CDX-IG9 & \multicolumn{2}{|c|}{Independent} & \multicolumn{2}{|c|}{Less Correlated} & \multicolumn{2}{|c|}{More Correlated}\\
\cline{2-7}
Tranches & LB & UB & LB & UB & LB & UB\\
\hline
0-3\% & 2.86\% & 3.82\% & 3.92\% & 4.15\% & 4.72\% & 4.88\%\\
3-7\% & 4.63\% & 6.17\% & 8.19\% & 9.02\% & 10.47\% & 11.17\%\\
7-10\% & 3.52\% & 4.44\% & 7.79\% & 8.22\% & 9.95\% & 10.40\%\\
10-15\% & 1.99\% & 2.13\% & 5.00\% & 5.02\% & 6.09\% & 6.11\%\\
15-30\% & 0.61\% & 0.66\% & 1.82\% & 1.88\% & 2.00\% & 2.07\%\\
30-60\% & 0.44\% & 0.48\% & 1.36\% & 1.44\% & 1.45\% & 1.56\%\\
60-100\% & 0.11\% & 0.12\% & 0.36\% & 0.38\% & 0.38\% & 0.40\%\\
\hline
\end{tabular}

\end{table}

\subsection{Single Default Event Trigger}
We take the 3Y to 5Y CDX-IG9 ATM tranche loss option as an example,
and consider an option that can be exercised at 3Y only if a trigger 
credit has defaulted before 3Y. We further assume that the trigger credit 
does not appear in the portfolio of the CDO tranche\footnote{If the name 
does appear in the tranche portfolio,
we can always replicate the original option by an equivalent tranche loss 
option without the trigger credit in the portfolio by adjusting the tranche 
attachment, detachment and the strike price of the option because the trigger 
credit has to be in the default state when the option has non-zero payoff.}. 
Because of the single credit trigger, the price bounds of this option cannot be 
obtained using a pure top-down model. 

We consider the price bound with a Maximum Entropy Markov chain under the 
\cite{self} model. Because of the conditional independence, whether the trigger name default
before 3Y does not change the distribution or the transition Markov Chain of the
common factor process; neither does it change the conditional default 
probabilities of any other names in the portfolio. Therefore, we can obtain 
the price bounds of this option 
by simply weighting the option payoff in \eqref{ub} and \eqref{lb} by the 3Y 
conditional default probability of the trigger credit:
\begin{align}
\label{lbs}
C^U &= \mathbb{E}[{\bf 1}_{\tau < t} \max(\sum_{t_i > t} d(0, t_i) c_i - K, 0)] \notag \\
		&= \mathbb{E}[\mathbb{E}[{\bf 1}_{\tau < t} \max(\sum_{t_i > t} d(0, t_i) c_i - K, 0)|X_t]] &:& \mbox{ Iterative expectation} \notag \\
		&= \mathbb{E}[\mathbb{E}[{\bf 1}_{\tau < t}|X_t] \mathbb{E}[\max(\sum_{t_i > t} d(0, t_i) c_i - K, 0)|X_t]] &:& \mbox { Conditional Independence} \notag \\
		&= \mathbb{E}[q(X_t, t) \mathbb{E}[\max(\sum_{t_i > t} d(0, t_i) c_i - K, 0)|X_t]]
\end{align}
Where $X_t$ is the value of the common factor at time $t$, $q(X_t, t)$ is the 
conditional default probability of the trigger credit. 
Similarly, we can get the following expression for the lower bound:
\begin{equation} 
\label{ubs}
C^L =  \mathbb{E}[q(X_t,t) \mathbb{E}[\max(\mathbb{E}[ \sum_{t_i > t} d(0, t_i) c_i | \mathcal{Y}_t] - K, 0)|X_t]] 
\end{equation}
Both of the bounds in \eqref{lbs} and \eqref{ubs} are easy to compute 
semi-analytically.

As shown in Table \ref{sntrig}, we computed the price bounds with two 
different default probabilities for the trigger credit, 5\% and 30\%. We 
also computed the price bounds with different correlations between the 
trigger credit default and the common market factor. As expected, the 
option is more valuable if the trigger credit is more risky. Table \ref{sntrig} 
also showed that the option is much more valuable if the trigger credit 
is more correlated to the common market factor. In the context of 
counterparty risk, this results in the so called ``wrong-way'' risk of 
buying tranche protection from a risky counterparty, i.e., the tranche 
protection worth much less if the counterparty is more likely to default 
when the portfolio suffers more losses.

The price bounds in Table \ref{sntrig} are much narrower than the 
comparable bounds from Maximum Entropy Markov chain in table \ref{upper} 
and \ref{bd1f}. The price bounds are very narrow even when the trigger name
has significant default probability; therefore there is clearly no 
need to build the full dynamic spread models for the single default
event trigger. Counterparty risk of tranches therefore can be very 
effectively priced and managed using this methodology.

In this example, we assume the option is exercised at 3Y if the trigger 
name defaults before 3Y, we refer to it as the ``exercise-at-maturity'' option. 
In a more realistic setting, the option holder has to exercise the option 
immediately if the trigger credit default, which is referred as  
``exercise-at-trigger''. The ``exercise-at-trigger'' option can be modeled as a 
series ``exercise-at-maturity'' options, with one option expires at every 
default observation date and is only exercisable if the trigger credit 
defaults between the previous default observation date and the current 
default observation date.  Given that he ``exercise-at-maturity'' option 
expires at a fixed maturity date and
the ``exercise-at-trigger'' option is a series of options that expires
at each default observation date from time 0 to the maturity, 
the ``exercise-at-trigger'' option's value is always less than the 
``exercise-at-maturity'' option with the same
maturity because an option is less valuable with shorter expiration.

\subsection{Generic Random Triggers \label{gtrigger}}
For more general triggers that involve multiple names, such as portfolio loss 
triggers or the 1st default event in a credit basket, we have to use Monte 
Carlo simulation of default time and recovery to compute the price 
bounds. The semi-analytical solutions for these complicated triggers often
gets too tedious comparing to the straight-forward Monte Carlo simulation.

The lower bound of the option depends on the term $\mathbb{E}[V_t | \mathcal{Y}_t]$
in \eqref{lb}. In section \ref{fixedtr}, we restricted ourselves to only condition 
on the common  factor $X_t$, which is the most convenient for semi-analytical solutions. 
However, in Monte Carlo simulation, we can easily add additional variables in the filtration
$\mathcal{Y}_t$ to the conditioning so that we can get better (or higher) lower bounds. The realized
portfolio loss $L_t$ is the next most useful factor to be included in the conditioning 
after the common factor $X_t$. Given the conditional independence, there is limited 
benefits to include individual names' default indicators in the conditioning after 
$X_t$ and $L_t$. If we only use $X_t, L_t$ as the two conditioning variables, 
the $\mathbb{E}[V_t | X_t, L_t]$ can be directly computed from the simulation by 
constructing a two dimensional grid that samples the $X_t$ and $L_t$ discretely. 
If there are more variables in the conditioning, we have to use the regression 
technique in the typical least square Monte Carlo methodology in \cite{ls}. 

Table \ref{mclb} shows the lower bound of the tranche loss option with deterministic
time trigger implied by the maximum entropy Markov chain conditioned on both $X_t$ 
and $L_t$. The results are obtained from a Monte Carlo simulation where half of
the simulated path is used to establish the $\mathbb{E}[V_t | X_t, L_t]$ by 
constructing a two-dimensional grid of $(X_t, L_t)$, and the other half of the 
simulated path is used to compute the actual lower bounds from the 
$\mathbb{E}[V_t | X_t, L_t]$.
Comparing with Table \ref{bd1f} and Table \ref{bd10y}, the lower bound for
junior tranches improved slightly by adding the realized loss $L_t$ in 
the conditioning. The lower bounds of senior tranches showed almost no 
improvements. 
\begin{table}
\caption{Max Entropy Lower Bounds Conditioned on $X_t$ and $L_t$ \label{mclb}}
\center

\footnotesize
\begin{tabular}{|c|rrr|rrr|}
\hline
CDX-IG9 & \multicolumn{3}{|c|}{3Y-5Y Lower Bounds} & \multicolumn{3}{|c|}{5Y-10Y Lower Bounds}\\
\cline{2-7}
Tranches & ITM & ATM & OTM & ITM & ATM & OTM\\
\hline
0-3\% & 41.39\% & 10.31\% & 0.00\% & 46.14\% & 6.25\% & 0.00\%\\
3-7\% & 26.74\% & 15.68\% & 2.72\% & 40.03\% & 18.14\% & 0.00\%\\
7-10\% & 14.45\% & 11.86\% & 8.08\% & 30.94\% & 17.27\% & 0.67\%\\
10-15\% & 7.23\% & 6.62\% & 5.54\% & 16.64\% & 12.38\% & 6.93\%\\
15-30\% & 2.12\% & 2.02\% & 1.88\% & 6.55\% & 5.95\% & 4.97\%\\
30-60\% & 1.50\% & 1.45\% & 1.34\% & 4.54\% & 4.22\% & 3.66\%\\
60-100\% & 0.39\% & 0.38\% & 0.35\% & 1.35\% & 1.26\% & 1.14\%\\
\hline
\end{tabular}
\end{table}

We also considered a more realistic example of callable tranche where 
a client sold protection to a bank on a senior 5Y IG9 tranche, and the 
client has the right to buy back the protection at the initial expected 
tranche loss if the IG9 portfolio loss is greater than a pre-determined 
threshold $\alpha$ at the 3Y. In this example, the trigger event and the option 
payoff are highly correlated as both of them are functions of the IG9 
portfolio loss, therefore, we cannot compute its price bounds by simply 
multiplying the tranche option payoffs in Table \ref{bd1f} by the probability of 
the trigger event. Instead we have to use the full Monte Carlo simulation to 
compute the price bounds, which are shown in Table \ref{lss}. The price 
bounds of the options to call tranche with portfolio loss triggers are 
also very tight. 

\begin{table}
\caption{Price Bounds of 3Y-5Y Option to Call Tranche \label{lss}}
\center

\footnotesize
\begin{tabular}{|c|rr|rr|rr|}
\hline
CDX-IG9 & \multicolumn{2}{|c|}{$\alpha$ = 4\%} & \multicolumn{2}{|c|}{$\alpha$ = 8\%} & \multicolumn{2}{|c|}{$\alpha$ = 12\%}\\
\cline{2-7}
Tranches & LB & UB & LB & UB & LB & UB\\
\hline
15-30\% & 2.02\% & 2.22\% & 1.85\% & 1.95\% & 1.08\% & 1.13\%\\
30-60\% & 1.45\% & 1.62\% & 1.36\% & 1.47\% & 0.95\% & 1.00\%\\
60-100\% & 0.38\% & 0.42\% & 0.36\% & 0.38\% & 0.26\% & 0.27\%\\
\hline
\end{tabular}
\end{table}

The price bounds of other types of options, such as gap risk and liquidation 
risk, can also be computed from Monte Carlo simulation of default times 
and recovery rate using similar methods as in the callable tranche.

\section{Conclusion}
In this study, we have shown that the tranche option prices can be effectively 
bounded from a default time copula. We argue that the default time copula
from the maximum entropy Markov chain is the most natural choice when we don't 
have relevant market observables for tranche options; we also argue that it is 
possible for a dealer to make market and dynamically hedge the senior tranche 
options solely based on their price bounds.

For the European tranche options and the tranche options with single name 
default triggers, both the upper bound and lower bound can be computed from 
semi-analytical methods without Monte Carlo simulation. The Greeks of the 
price bounds can be computed by perturbing the market inputs and re-valuing. 
When the pricing bounds of an option is narrow, a dealer 
can treat the true value the option as an average of the lower 
and upper bound, and dynamically hedge an European tranche option book by 
averaging the Greeks for the lower bound and upper bound. A senior tranche 
option book could be managed without much additional effort
than the effort required to risk-manage a typical index or bespoke tranche book.

The fact that the upper and lower bounds can be hedged using static 
instruments makes it possible to profit from the market mis-pricing of the tranche 
options. In the event that the market tranche option prices are out of the 
pricing bounds, the dealer can take the corresponding option position and 
a hedge position for the bound that the option value violates. For example,
if a European tranche call option on a $A$ to $D$ tranche with strike $K$
is priced higher in the market than its upper bound, then a dealer can 
sell the call option and hedge it by buying protection on a $A+K$ to $D$
tranche. This will results in a positive profit without any future risk, 
i.e., an arbitrage opportunity. However,
in reality, the $A+K$ to $D$ tranche may not be liquid itself, therefore,
the tranche option may have to be hedge dynamically using other liquid 
instruments. The dynamic hedging of the lower or upper bound of the tranche
options is no more complicated than the common practice of dynamically 
hedging the off-the-run index tranches or bespoke tranches. 
 
Contrary to a common stereotype, the tranche options prices are primarily 
determined by the $JDDT$ and default time copula. Other systemic and
idiosyncratic factors beyond the default time copula contribute a relatively
small portion of the pricing uncertainty. Using top-down 
models to price tranche options could result in mis-pricing given the 
top-down models ignores a large amount of static single name market 
information which is important in determining the tranche option prices. 

The methodology to obtain price bounds of tranche options described in 
this paper could also play an important role in the management of 
counterparty risk, gap risk and liquidation risk. As shown in section 
\ref{gtrigger}, the price bounds are often very narrow for tranche 
options with random trigger event, therefore, these practical problems 
can also be effectively addressed using the price bounds. 

\bibliographystyle{apsr}
\bibliography{../creditref}

\vspace{3cm}
\appendix
\Large {\bf Appendix} \normalsize

\section{Approximate the Tranche PV Option\label{a_tr}}
Consider a generic tranche with non-zero coupon, and whose protection payment 
is settled at the time of default. The MTM of such a tranche can be written as: 
\begin{equation*}
V_t = PROT_t - s PV01_t 
\end{equation*}
where $PROT$ is the protection PV and the $PV01$ is the PV of a 1bps coupon 
payment, and $s$ is the contractual coupon. Since the main risk factor of a 
tranche is its terminal loss, we can view the $PROT_t$ and $PV01_t$ as functions 
of the expected terminal tranche loss $l_t = \mathbb{E}[L_T(A, D)|\mathcal{F}_t]$. 
Therefore, we can expand the $PROT(l_t)$ and $PV01(l_t)$ around the current 
tranche expected loss $l_t^0 = \mathbb{E}[L_T(A, D)|\mathcal{F}_0]$:
\begin{align*}
PROT(l_t) &\approx PROT(l_t^0) + \frac{\partial PROT(l_t^0)}{\partial l_t} (l_t - l_t^0) \\
PV01(l_t) &\approx PV01(l_t^0) + \frac{\partial PV01(l_t^0)}{\partial l_t}(l_t-l_t^0)
\end{align*}
Therefore, we have:
\begin{align*}
V_t &\approx PROT(l_t^0) - s PV01(l_t^0) + ( \frac{\partial PROT(l_t^0)}{\partial l_t} - s \frac{\partial PV01(l_t^0)}{\partial l_t}) (l_t - l_t^0) \\
	&= a + b l_t
\end{align*}
Where $a, b$ are constants which are obvious from the above equation. This 
effectively approximates the MTM of a regular tranche by a linear function 
of $l_t$. The first order derivatives can be obtained 
from the default time copula and the price of an option on tranche PV 
with strike $K$ can be approximated by a tranche loss option:
\begin{align*}
C &= \mathbb{E}[ d(0, t) {\bf 1}_{\tau = t} \max(V_t - K, 0) ] \\
	&\approx \mathbb{E}[ d(0, t) {\bf 1}_{\tau = t} \max(a + b l_t - K, 0) ] \\
	&= \mathbb{E}[ d(0, t) {\bf 1}_{\tau = t} b \max(l_t - \frac{K - a}{b}, 0) ]
\end{align*}
Therefore, the tranche loss option bounds discussed in this paper can be applied 
to produce the bounds for the tranche PV option. In practice, the tranche 
PV is mainly driven by its terminal loss, thus the approximation should 
be adequate in most situations. 

\section{Finding the Lowest Lower Bound \label{a_opt}}

We want to find the lowest lower bound among all possible Markov chains 
from two loss (or common factor) distributions; this problem can be formulated 
as an nonlinear optimization with linear constraints. Here we use the LLB for 
the loss distribution to describe the optimization setup, the LLB of the common 
factor process can be solved using the exact same method. 

Suppose the portfolio loss distribution at two time horizons $t<T$ are sampled 
by a discrete loss grid $l_i$, the corresponding loss probability are given by 
$p_i(t)$ and $p_i(T)$, which has to satisfy the usual constraints of being valid
loss distributions.

A discrete Markov chain is parameterized by its transition probability $q_{ij} 
= \mathbb{P} \{ l_j | l_i\}$, obviously $q_{ij}$ is zero if $i>j$. The transition
probability has to satisfy the following constraints from the initial 
loss distribution:
\begin{align}
\label{lincon}
\sum_i q_{ij} &= p_i(t) \nonumber \\
\sum_j q_{ij} &= p_j(T)
\end{align}
The lower bound from the Markov chain $q_{ij}$ is a nonlinear objective function 
that we have to minimize by adjusting the $q_{ij}$. Therefore, this is a nonlinear
optimization problem with linear constraints given in (\ref{lincon}). The dimension
of $q_{i,j}$ is about $N^2/2$ where $N$ is the number of discrete samples on the
loss distribution. For a $N$ less than 40, this problem can be solved 
in a few minutes using Powell's TOLMIN algorithm.

\end{document}